\documentclass[aps,showpacs,preprintnumbers,nofootinbib]{revtex4}
\usepackage{amssymb}
\usepackage{amsfonts}
\usepackage{amsmath}
\usepackage{graphicx}
\usepackage{dcolumn}
\usepackage{bm}
\usepackage[latin1]{inputenc}
\usepackage[dvips]{hyperref}
\usepackage{graphicx}

\def\be{\begin{equation}}
\def\ee{\end{equation}}
\def\bea{\begin{eqnarray}}
\def\eea{\end{eqnarray}}

\begin{document}

\title{Quasinormal modes and greybody factors of a four-dimensional Lifshitz black hole with $z=0$} 
\author{Marcela Catal\'{a}n}
\email{marceicha@gmail.com}
\affiliation{Departamento de Ciencias F\'{\i}sicas, Facultad de Ingenier\'{\i}a y Ciencias, Universidad de La Frontera, Avenida Francisco Salazar
01145, Casilla 54-D, Temuco, Chile.}
\author{Eduardo Cisternas}
\email{eduardo.cisternas@ufrontera.cl}
\affiliation{Departamento de Ciencias F\'{\i}sicas, Facultad de Ingenier\'{\i}a y Ciencias, Universidad de La Frontera, Avenida Francisco Salazar
01145, Casilla 54-D, Temuco, Chile.}
\author{P.~A.~Gonz\'{a}lez}
\email{pablo.gonzalez@udp.cl}
\affiliation{Facultad de Ingenier\'{\i}a, Universidad Diego Portales, Avenida Ej\'{e}%
rcito Libertador 441, Casilla 298-V, Santiago, Chile.}
\author{Yerko V\'{a}squez}
\email{yvasquez@userena.cl}
\affiliation{Departamento de F\'{\i}sica, Facultad de Ciencias, Universidad de La Serena,\\ 
Avenida Cisternas 1200, La Serena, Chile.}
\date{\today }

\begin{abstract}

We study scalar perturbations for a
four-dimensional asymptotically Lifshitz black hole in conformal gravity with dynamical exponent $z=0$, and spherical topology for the
transverse section, and we find analytically and numerically the quasinormal
modes for scalar fields for some special cases.
Then, we study the stability of these black holes under scalar field perturbations and the greybody factors.

\end{abstract}

\maketitle


\section{Introduction}

Lifshitz spacetimes have received considerable attention from the condensed matter point of view due to the AdS/CFT correspondence, i.e., searching for gravity duals of Lifshitz fixed points for condensed matter physics and quantum chromodynamics \cite{Kachru:2008yh}. From the quantum field theory point of view, there are many invariant scale theories of interest
when studying such critical points. Such theories exhibit the anisotropic
scale invariance $t\rightarrow \chi ^{z}t$, $x\rightarrow \chi x$, with $z\neq
1$, where $z$ is the relative scale dimension of time and space, and these
are of particular interest in studies of critical exponent theory and phase
transitions. Systems with such behavior appear, for instance, in the
description of strongly correlated electrons. The importance of possessing a
tool to study strongly correlated condensed matter systems is beyond
question, and consequently much attention has focused on this area in
recent years.

One of the most well studied systems in the context of gauge/gravity duality, is the holographic superconductor. In its simplest form, the gravity sector is a gravitating system with a cosmological constant, a gauge field and a charged scalar field with a potential. The dynamics of the system defines a critical temperature above which the system finds itself in its normal phase and the scalar field does not have any dynamics. Below the critical temperature the system undergoes a phase transition to a new configuration. 
From the gravity side this is interpretated as the black hole to acquire hair while from boundary conformal field theory site this is interpretated as a condensation of the scalar field and the system enters a superconducting phase. In this sense, Lifshitz holographic superconductivity has been a topic of numerous studies and interesting properties are found when one generalizes the gauge/gravity duality to non-relativistic situations \cite{Hartnoll:2009ns, Brynjolfsson:2009ct,  Sin:2009wi, Schaposnik:2012cr, Momeni:2012tw, Bu:2012zzb, Keranen:2012mx, Zhao:2013pva, Lu:2013tza, Tallarita:2014bga}. 

The Lifshitz spacetimes are described by the metrics 
\begin{equation}
ds^2=- \frac{r^{2z}}{\ell^{2z}}dt^2+\frac{\ell^2}{r^2}dr^2+\frac{r^2}{\ell^2} d\vec{x}%
^2~,  \label{lif1}
\end{equation}
where $\vec{x}$ represents a $D-2$ dimensional spatial vector, $D$ is the
spacetime dimension and $\ell$ denotes the length scale in the geometry. If $z=1$, the spacetime is the
usual anti-de Sitter metric in Poincar\'{e} coordinates. Furthermore, all scalar
curvature invariants are constant and these spacetimes have a null curvature
singularity at $r\rightarrow 0$ for $z\neq 1$, which can be seen by
computing the tidal forces between infalling particles. This singularity is
reached in finite proper time by infalling observers, so the spacetime is
geodesically incomplete \cite{Horowitz:2011gh}. The metrics of Lifshitz
black holes asymptotically have the form (\ref{lif1}); however, obtaining
analytical solutions does not seem to be a trivial task, and therefore
constructing finite temperature gravity duals requires the introduction of
strange matter content with a theoretical motivation that is not clear.
Another way of finding such a Lifshitz black hole solution is by considering
carefully-tuned higher-curvature modifications to the Hilbert-Einstein
action, as in new massive gravity (NMG) in 3-dimensions or $R^2$ corrections
to general relativity. This has been done, for instance, in \cite%
{AyonBeato:2009nh, Cai:2009ac, AyonBeato:2010tm, Dehghani:2010kd}. A
4-dimensional topological black hole with $z=2$ was
found in \cite{Mann:2009yx, Balasubramanian:2009rx} and a set of analytic Lifshitz black holes in
higher dimensions for arbitrary $z$ in \cite{Bertoldi:2009vn}. Lifshitz black holes with arbitrary dynamical exponent in Horndeski theory were found in \cite{Bravo-Gaete:2013dca} and non-linearly charged Lifshitz black holes for any exponent $z>1$ in \cite{Alvarez:2014pra}. Thermodynamically, it is difficult to compute conserved
quantities for Lifshitz black holes; however, progress was made on the
computation of mass and related thermodynamic quantities by using the ADT
method \cite{Devecioglu:2010sf, Devecioglu:2011yi} as well as the Euclidean
action approach \cite{Gonzalez:2011nz, Myung:2012cb}. Also, phase
transitions between Lifshitz black holes and other configurations with
different asymptotes have been studied in \cite{Myung:2012xc}. However, due
to their different asymptotes these phases transitions do not occur. 

Conformal gravity is a four-derivative theory and is perturbatively renormalizable \cite{Stelle:1976gc, Stelle:1977ry}. Also, it contains ghost-like modes in the form of massive spin-2 excitations. However, a solution to the ghost problem in fourth order derivative theories was shown in  \cite{Mannheim:2006rd} by using the method of Dirac constraints \cite{Dirac} to quantize the Pais-Uhlenbeck fourth order oscillator model \cite{Pais:1950za}.
In this work, we consider a matter distribution outside the event horizon of
the Lifshitz black hole in $4$-dimensions in conformal gravity with a spherical transverse section and dynamical exponent $z=0$. It is worth mentioning that for $z=0$ the previously mentioned anisotropic scale invariance corresponds to space-like scale invariance with no transformation of time. The matter is parameterized by scalar fields minimally and conformally coupled to gravity. Then, we obtain
analytically and numerically the quasinormal frequencies (QNFs) \cite{Regge:1957td, Zerilli:1971wd,
Zerilli:1970se, Kokkotas:1999bd, Nollert:1999ji, Konoplya:2011qq} for scalar fields, after which we study
their stability under scalar perturbations. Also, we compute the reflection and transmission coefficients and the absorption cross section.

The study of the QNFs gives
information about the stability of black holes under matter fields that
evolve perturbatively in their exterior region, without backreacting on the
metric. In general, the oscillation frequencies are complex, where the real
part represents the oscillation frequency and the imaginary part describes
the rate at which this oscillation is damped, with the stability of the
black hole being guaranteed if the imaginary part is negative. The
QNFs are independent of the initial
conditions and depend only on the parameters of the black hole (mass,
charge and angular momentum) and the fundamental constants (Newton constant
and cosmological constant) that describe a black hole, just like the
parameters that define the test field.  On the other hand, the QNFs determine how fast a thermal state in the
boundary theory will reach thermal equilibrium according to the AdS/CFT
correspondence \cite{Maldacena:1997re}, where the relaxation time of a
thermal state is proportional to the inverse
of the imaginary part of the QNFs of the dual gravity background, which was
established due to the QNFs of the black hole being related to the poles of
the retarded correlation function of the corresponding perturbations of the
dual conformal field theory \cite{Birmingham:2001pj}. Fermions on a Lifshitz
background were studied in \cite{Alishahiha:2012nm} by using the
fermionic Green's function in 4-dimensional Lifshitz spacetime with $z=2$; the authors considered a non-relativistic (mixed) boundary condition
for fermions and showed that the spectrum has a flat band. Also, the Dirac quasinormal modes (QNMs) for a 4-dimensional Lifshitz black hole were studied in \cite{Catalan:2013eza}. Generally, the Lifshitz black holes are stable under scalar perturbations, and the QNFs show the absence
of a real part \cite{CuadrosMelgar:2011up, Gonzalez:2012de, Gonzalez:2012xc,
Myung:2012cb, Becar:2012bj,Giacomini:2012hg}. The QNFs have been calculated by means of numerical and analytical techniques, some remarkably numerical methods are: the Mashhoon method, Chandrasekhar-Detweiler, WKB method, Frobenius method, method of continued fractions, Nollert, asymptotic iteration method (AIM) and improved AIM among others. In the context of black hole
thermodynamics, QNMs allow the quantum area spectrum of the black hole
horizon to be studied \cite{CuadrosMelgar:2011up} as well as the mass and
the entropy spectrum. 

On the other hand, knowledge of black holes perturbations is also useful for studying the Hawking radiation, which is a semiclassical effect and gives the thermal radiation emitted by a black hole. At the event horizon, the Hawking radiation is in fact blackbody radiation. However, this radiation still has to traverse a non-trivial curved spacetime geometry before reaching a distant observer that can detect it. The surrounding spacetime thus works as a potential barrier for the radiation, giving a deviation from the blackbody radiation spectrum, seen by an asymptotic observer \cite{Maldacena:1996ix}. Thus the total flux observed at infinity is that of a $D$-dimensional greybody at the Hawking temperature. The factors that modify the spectrum emitted by a black hole are known as greybody factors and can be obtained through the classical scattering (for a review see \cite{Harmark:2007jy}). In this sense, the scalar greybody factors for an asymptotically Lifshitz black hole were studied in \cite{Gonzalez:2012xc, Lepe:2012zf}, and particle motion on these geometries in \cite{Olivares:2013zta, Olivares:2013uha, Villanueva:2013gra}.

The paper is organized as follows. In Sec. II we give a brief review of
the $4$-dimensional Lifshitz black hole in conformal gravity. In Sec. III
we calculate the QNFs of scalar perturbations for the $4$-dimensional Lifshitz
black hole with spherical topology and $z=0$ for some special cases analytically and numerically by using the improved AIM. Then, in Sec. IV, we study the reflection and transmission coefficients and the absorption cross section. Finally, our conclusions are
in Sec. V.

\section{$4$-dimensional asymptotically Lifshitz black hole in conformal gravity}

In this work we consider a matter distribution described by a scalar field outside the event horizon of a four-dimensional asymptotically Lifshitz black hole in conformal gravity with $z=0$ and spherical topology for the transverse section \cite{Lu:2012xu}. Conformal gravity is a limit case of Einstein-Weyl gravity. The action of Einstein-Weyl gravity is given by
\begin{equation}
S=\frac{1}{2k^{2}}\int {\sqrt{-g}d^{4}x}\left( {\mathcal{R}-2\Lambda +\frac{1}{2}%
\alpha |Weyl|^{2}}\right) ~,
\end{equation}%
where 
\begin{equation}
|Weyl|^{2}=\mathcal{R}^{\mu \nu \rho \sigma }\mathcal{R}_{\mu \nu \rho \sigma }-2\mathcal{R}^{\mu \nu
}\mathcal{R}_{\mu \nu }+\frac{1}{3}\mathcal{R}^{2}~,
\end{equation}%
$\mathcal{R}$ is the Ricci scalar and $\Lambda $ is the cosmological constant. When $\alpha$ goes to infinity we have the special case of conformal gravity, and the field equations in vacuum are given by $B_{\mu \nu}=0$, where $B_{\mu \nu}$ is the Bach tensor defined by:
\begin{equation}
B_{\mu \nu}=(\nabla^{\rho}\nabla^{\sigma}+\frac{1}{2}\mathcal{R}^{\rho \sigma})C_{\mu \nu \rho \sigma}~,
\end{equation}
where $C_{\mu \nu \rho \sigma}$ is the Weyl tensor. The following metric solves the field equations \cite{Lu:2012xu}
\begin{eqnarray}
ds^{2} &=&-fdt^{2}+\frac{4\ell^2dr^{2}}{r^{2}f}+r^{2}d\Omega _{2,k}^{2}~,  \label{metrica}
\\
f &=&1+\frac{\lambda }{r^{2}}+\frac{\lambda ^{2}-k^{2}\ell^4}{3r^{4}}~.
\end{eqnarray}%
For $k=\pm 1$, there is an event horizon at the largest root of $f$, given by 
\begin{equation}
r_{+}^{2}=\frac{1}{6}\left( \sqrt{3(4\ell^4-\lambda ^{2})}-3\lambda \right) ~,
\end{equation}%
and for $k=0$ the singularity is naked. 
Note that the requirement $%
r_{+}^{2}>0$ implies that $-2\ell^2\leq \lambda <\ell^2$. When $\lambda=-2\ell^2$ the solution becomes extremal, and for $k=1$ the entropy vanishes in this case. The Kretschmann scalar (for $k=1$) is given by
\begin{equation}
\mathcal{R}_{\mu \nu \rho \sigma} \mathcal{R}^{\mu \nu \rho \sigma}=\frac{9r^8+6(\lambda -4\ell^2)r^6+(50\ell^4+\lambda (19 \lambda -24\ell^2))r^4+2(\lambda ^2-\ell^4)(21 \lambda-4\ell^2)r^2+25(\lambda ^2-\ell^4)^2}{12\ell^4r^8}~,
\end{equation}
therefore, there is a curvature singularity at $r=0$. In the next section, we determine the QNFs by considering the Klein-Gordon equation in this background and by establishing the boundary
conditions on the scalar field at the horizon and at spatial infinity.

\section{Quasinormal modes of a $4$-dimensional Lifshitz black hole}

The QNMs of scalar perturbations in the background of a four-dimensional asymptotically Lifshitz black hole in conformal gravity with dynamical exponent $z=0$ are given by the scalar field solution of Klein-Gordon equation with suitable boundary conditions. This means there are only ingoing waves on the event horizon and we consider that the scalar field vanishes at spatial infinity, known as Dirichlet boundary conditions. These fields are considered as mere test fields, without backreaction over the spacetime itself. Therefore, it is not necessary for such fields to have the same symmetries as the background spacetime. On the other hand, if one considers the backreaction of the matter fields over the spacetime, in order to look for exact solutions to the field equations, the relation between symmetries of the spacetime and the matter fields is not trivial, for a recent study about symmetry inheritance of scalar fields see \cite{Smolic:2015txa} and references therein. In the case considered here, the gravitational field equations imply that the trace of the stress-energy tensor must vanish, due to the Bach tensor is traceless, therefore if one go beyond the probe-field approximation, this implies that the stress-energy tensor of the matter fields must be traceless. Based on these arguments, first we will consider a test scalar field minimally coupled to curvature, then we will consider a test scalar field conformally coupled to curvature, which have a traceless stress-energy tensor, and we find analytically and numerically the quasinormal frequencies for scalar fields for some special cases.

\subsection{Scalar field minimally coupled  to gravity}
\begin{figure}[h]
\begin{center}
\includegraphics[width=0.7\textwidth]{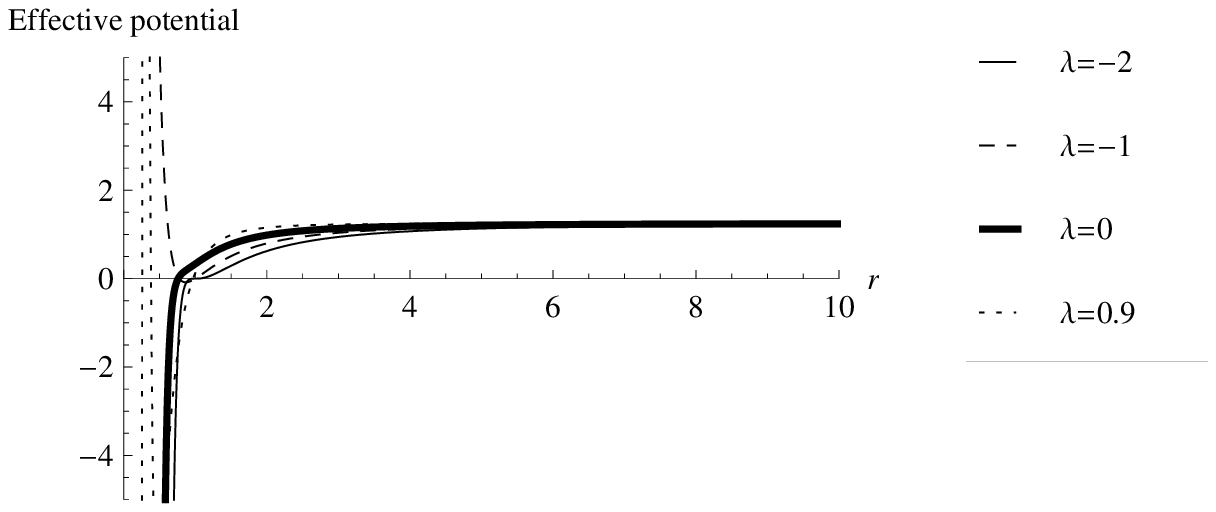}
\end{center}
\caption{The effective potential as a function of $r$, for $m=1$, $\ell=1$ and $\kappa=2$.} \label{potential1}
\end{figure}

\begin{figure}[h]
\begin{center}
\includegraphics[width=0.7\textwidth]{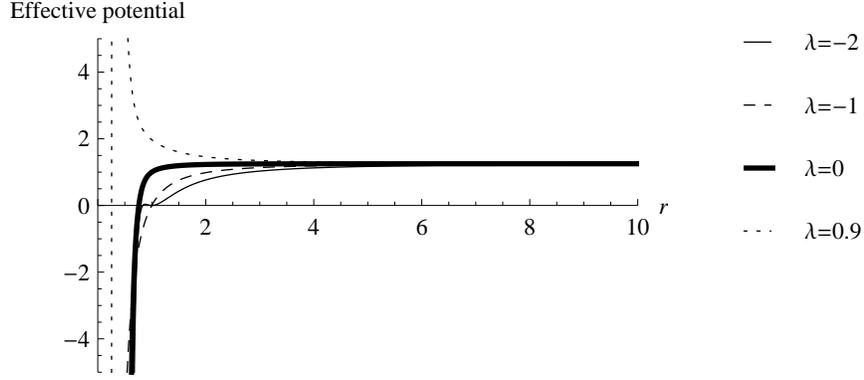}
\end{center}
\caption{The effective potential as a function of $r$, for $m=1$, $\ell=1$ and $\kappa=0$.} \label{potential0}
\end{figure}

In this section we calculate the QNMs of the $z=0$ Lifshitz black hole for a test scalar field minimally coupled to gravity. The Klein-Gordon equation in curved spacetime is 
\begin{equation}
\frac{1}{\sqrt{-g}}\partial _{\mu }\left( \sqrt{-g}g^{\mu \nu }\partial
_{\nu }\right) \psi =m^{2}\psi ~,  \label{KG}
\end{equation}%
where $m$ is the mass of the scalar field $\psi $, which is minimally
coupled to curvature. By means of the following ansatz 
\begin{equation}
\psi =e^{-i\omega t}R(r)Y(\theta ,\phi )~,
\end{equation}%
where $Y(\theta ,\phi )$ is a normalizable harmonic function on the two-sphere
which satisfies 
\begin{equation}\label{nabla}
\nabla ^{2}Y=-\kappa Y~, 
\end{equation}
being $\kappa =l\left( l+1\right)$ the eigenvalues for the spheric manifold, with $l=0,1,2,...$~, the Klein-Gordon equation reduces to
\begin{equation}
\frac{1}{4r}\partial _{r}\left( r^{3}f\left( r\right) \partial _{r}R\right)
+\left( \frac{\omega ^{2} \ell^2}{f\left( r\right) }-\frac{\kappa \ell^2 }{r^{2}}%
-m^{2} \ell^2\right) R\left( r\right) =0~.  \label{first}
\end{equation}%
Now, by considering $R(r)=K(r)/r$ and by introducing the tortoise coordinate 
$r_{\ast }$, given by $dr_{\ast }=\frac{2 \ell dr}{rf(r)}$, the latter equation
can be rewritten as a one-dimensional Schr\"{o}dinger equation 
\begin{equation}
\left[ \partial _{r_{\ast }}^{2}+\omega ^{2}-V_{eff}(r)\right] K(r_{\ast
})=0~,
\end{equation}%
where the effective potential is given by 
\begin{equation}
V_{eff}(r)=\frac{f(r)}{4}\left[ \frac{f(r)}{\ell^2}+\frac{rf^{\prime }(r)}{\ell^2}+\frac{4\kappa }{r^{2}}%
+4m^{2}\right] ~.
\end{equation}%
In Fig.~(\ref{potential1}) we plot the effective potential for $\kappa=2$ and in Fig.~(\ref{potential0}) for $\kappa=0$ and different values of the parameter $\lambda$. Note that when $r\rightarrow \infty$ the effective potential goes to $1/(4\ell^2)+m^2$.

\subsubsection{Case $\kappa=0$}
\begin{figure}[h]
\begin{center}
\includegraphics[width=0.6\textwidth]{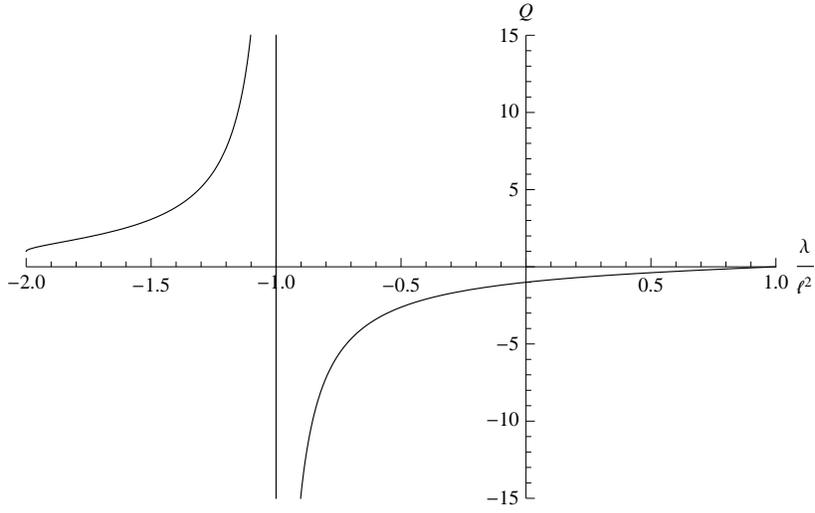}
\end{center}
\caption{$Q$ as a function of $\lambda / \ell^2$.} \label{Q}
\end{figure}

In order to find analytical solutions to the radial equation (\ref%
{first}), we perform the change of variables $y=r^{2}$ and get the
following equation:
\begin{equation}
\label{13}
y\left( y-y_{+}\right) \left( y-y_{-}\right) R^{\prime \prime }\left(
y\right) +y\left( 2y+\lambda \right) R^{\prime }\left( y\right) +\left( 
\frac{\omega ^{2} \ell^2y^{3}}{\left( y-y_{+}\right) \left( y-y_{-}\right) }-\kappa \ell^2
-m^{2} \ell^2 y\right) R\left( y\right) =0~,
\end{equation}
where the prime denotes the derivative with respect to $y$, and $y_{+}$ and $%
y_{-}$ are the roots of 
\begin{equation}
f\left( y\right) =1+\frac{\lambda }{y}+\frac{\lambda ^{2}-\ell^4}{3y^{2}}~,
\end{equation}
and are given by
\begin{equation}
y_{\pm }=-\frac{\lambda }{2}\pm \sqrt{-\frac{\lambda ^{2}}{12}+\frac{\ell^4}{3}}~.
\end{equation}
Additionally, performing  another change of variable $z=1-\frac{y_{+}}{y}$
and noting that $\lambda =-\left( y_{+}+y_{-}\right) $, we arrive at the
following expression
\begin{equation}
z\left( z-1\right) \left( z-\left( 1-Q\right) \right) R^{\prime \prime
}\left( z\right) -\left( z-1\right) \left( 1-Q-2z\right) R^{\prime }\left(
z\right) +\left( \frac{\omega ^{2} \ell^2Q^{2}}{z\left( z-1\right) \left( z-\left(
1-Q\right) \right) }+\frac{\kappa \ell^2 }{y_{-}}-\frac{m^{2} \ell^2 Q}{z-1}\right) R\left(
z\right) =0~,  \label{rad}
\end{equation}
where we have defined $Q=y_{+}/y_{-}$ and now a prime means derivative with
respect to $z$. In Fig.~(\ref{Q}), we plot $Q$ as a function of $\lambda / \ell^2$ and we observe that $Q$ can be positive or negative depending on the values of $\ $the
parameter $\lambda $: 
\begin{eqnarray}
Q &>&1\text{ for }-2\ell^2\leq \lambda <-\ell^2~, \\
Q &<&0\text{ for }-\ell^2<\lambda <\ell^2~.
\end{eqnarray}
On the other hand, equation (\ref{rad}) can be manipulated and put into the following form
\begin{gather}
R^{\prime \prime }\left( z\right) +\left( \frac{1}{z}+%
\frac{1}{z-\left( 1-Q\right) }\right) R^{\prime }\left( z\right) +
\label{ecradial} \\
\left( \frac{\omega ^{2} \ell^2Q^{2}/(1-Q)}{z}+\frac{Q\left( \omega
^{2} \ell^2-m^{2}\ell^2\right) }{z-1}+\frac{\kappa \ell^2 }{y_{-}}-\frac{Q\omega ^{2} \ell^2/(1-Q)}{%
z-\left( 1-Q\right) }\right) \frac{1}{z\left( z-1\right) \left( z-\left(
1-Q\right) \right) }R\left( z\right) =0~.  \notag
\end{gather}
We note that for $\kappa =0$ this equation corresponds to a Riemann
differential equation, whose general form is \cite{M. Abramowitz}
\begin{eqnarray}
\nonumber&&\frac{d^{2}w}{dz^{2}}+\left( \frac{1-\alpha -\alpha ^{\prime }}{z-a}+\frac{%
1-\beta -\beta ^{\prime }}{z-b}+\frac{1-\gamma -\gamma ^{\prime }}{z-c}%
\right) \frac{dw}{dz}+ \\
&&\left( \frac{\alpha \alpha ^{\prime }\left( a-b\right) \left( a-c\right) }{%
z-a}+\frac{\beta \beta ^{\prime }\left( b-c\right) \left( b-a\right) }{z-b}+%
\frac{\gamma \gamma ^{\prime }\left( c-a\right) \left( c-b\right) }{z-c}%
\right) \frac{w}{\left( z-a\right) \left( z-b\right) \left( z-c\right) }=0~,
\end{eqnarray}
where $a,b$ and $c$ are the singular points, and the exponents $\alpha ,\alpha
^{\prime },\beta ,\beta ^{\prime },\gamma$ and $\gamma ^{\prime }$ are subject to
the condition
\begin{equation}
\alpha +\alpha ^{\prime }+\beta +\beta ^{\prime }+\gamma +\gamma ^{\prime }=1~.
\end{equation}
The complete solution is denoted by the symbol
\begin{equation}
w=P\left\{ 
\begin{array}{cccc}
a & b & c &  \\ 
\alpha  & \beta  & \gamma  & z \\ 
\alpha ^{\prime } & \beta ^{\prime } & \gamma ^{\prime } & 
\end{array}%
\right\}~,
\end{equation}
and the Riemann $P$ function can be reduced to the hypergeometric function
through
\begin{equation}
w=\left( \frac{z-a}{z-b}\right) ^{\alpha }\left( \frac{z-c}{z-b}\right)
^{\gamma }P\left\{ 
\begin{array}{cccc}
0 & \infty & 1 &  \\ 
0 & \alpha +\beta +\gamma & 0 & \frac{\left( z-a\right) \left( c-b\right) }{%
\left( z-b\right) \left( c-a\right) } \\ 
\alpha ^{\prime }-\alpha & \alpha +\beta ^{\prime }+\gamma & \gamma ^{\prime
}-\gamma & 
\end{array}%
\right\}~,
\end{equation}
where the $P$ function is now Gauss hypergeometric function \cite{M. Abramowitz}.
We observe that, in the radial equation (\ref{ecradial}), the regular
singular points $a,b$ and $c$ have the values
\begin{equation}
a=0~,\text{ \ }b=1-Q~,\text{ \ }c=1~,
\end{equation}
and the exponents are given by
\begin{eqnarray}
\alpha &=&\pm \frac{i\omega \ell y_+}{y_+-y_-}~, \alpha ^{\prime } =\mp \frac{i\omega \ell y_+}{y_+-y_-}~, \\
\beta &=&\pm \frac{i\omega \ell \left|y_-\right|}{y_+-y_- }~, \beta ^{\prime } =\mp \frac{i\omega \ell \left|y_-\right|}{y_+-y_- }~, \\
\gamma & = & \frac{1}{2}\pm \sqrt{ \frac{1}{4} - \left(\omega ^{2}-m^{2} \right) \ell^2 }~, \gamma ^{\prime } =\frac{1}{2}\mp \sqrt{ \frac{1}{4} -\left(\omega ^{2}-m^{2}\right) \ell^2 }~.
\end{eqnarray}
Therefore, the solution to equation (\ref{ecradial}) can be written as
\begin{eqnarray}
\nonumber R\left( z\right) &=&C_{1}\left( \frac{z}{z-1+Q}\right) ^{\alpha }\left( 
\frac{z-1}{z-1+Q}\right) ^{\gamma }{_2}F{_1}\left( A,B,C,\frac{Qz}{z-(1-Q)}\right) +
\\
&&C_{2}\left( \frac{z}{z-1+Q}\right) ^{\alpha ^{\prime }}\left( \frac{z-1}{%
z-1+Q}\right) ^{\gamma }{_2}F{_1}\left( A-C+1,B-C+1,2-C,\frac{Qz}{z-(1-Q)}\right)~,
\end{eqnarray}
where we have defined the constants $A,B$ and $C$ as
\begin{eqnarray}
\nonumber A &=&\alpha +\beta +\gamma~, \\
\nonumber  B &=&\alpha +\beta ^{\prime }+\gamma~, \\
C &=&1+\alpha -\alpha ^{\prime }~.
\end{eqnarray}
In the near-horizon limit, the above expression behaves as
\begin{equation} \label{hh}
R\left( z\rightarrow 0\right) =\frac{(-1)^\gamma C_{1}}{(-1+Q)^{\alpha+\gamma}}z^{\alpha }+\frac{(-1)^\gamma C_{2}}{(-1+Q)^{\alpha^{\prime}+\gamma}}z^{\alpha ^{\prime }}~,
\end{equation}
Now, we impose as a boundary condition that classically nothing can escape
from the event horizon. So, choosing the exponent $\alpha $ as
\begin{equation}
\alpha =-\frac{i\omega \ell y_+}{y_+-y_-}~,
\end{equation}
implies that we must take $C_{2}=0$ in order to have only ingoing waves at
the horizon. Therefore, our solution simplifies to
\begin{equation}
\label{horizonk0}
R\left( z\right) =C_{1}\left( \frac{z}{z-1+Q}\right) ^{\alpha }\left( \frac{%
z-1}{z-1+Q}\right) ^{\gamma }{_2}F{_1}\left( A,B,C,\frac{Qz}{z-(1-Q)}\right)~.
\end{equation}
Now, we implement boundary conditions at spatial infinity. In order to do so
we employ the Kummer relations \cite{M. Abramowitz}, and write the solution as
\begin{eqnarray}
\nonumber R\left( z\right)  &=&C_{1}\left( \frac{z}{z-1+Q}\right) ^{\alpha }\left( 
\frac{z-1}{z-1+Q}\right) ^{\gamma }\frac{\Gamma \left( C\right) \Gamma
\left( C-A-B\right) }{\Gamma \left( C-A\right) \Gamma \left( C-B\right) }%
{_2}F{_1}\left( A,B,A+B-C,1-\frac{Qz}{z-\left( 1-Q\right) }\right) + \\
\nonumber &&C_{1}\left( 1-Q\right) ^{\gamma ^{\prime }-\gamma }\left( \frac{z}{z-1+Q}%
\right) ^{\alpha }\left( \frac{z-1}{z-1+Q}\right) ^{\gamma ^{\prime }}\frac{%
\Gamma \left( C\right) \Gamma \left( A+B-C\right) }{\Gamma \left( A\right)
\Gamma \left( B\right) }\times  \\
&&{_2}F{_1}\left( C-A,C-B,C-A-B+1,1-\frac{Qz}{z-\left( 1-Q\right) }\right)~.
\end{eqnarray}
At the limit $z\rightarrow 1$, the above solution becomes
\begin{equation}
R\left( z\rightarrow 1\right) =\frac{C_{1}}{Q^{\alpha+\gamma}}\left( z-1\right) ^{\gamma }
\frac{\Gamma \left( C\right) \Gamma \left( C-A-B\right) }{\Gamma \left(
C-A\right) \Gamma \left( C-B\right) }+\frac{C_{1}\left( 1-Q\right)^{\gamma^{\prime }-\gamma }}{Q^{\alpha+\gamma^{\prime }}}\left( z-1\right) ^{\gamma ^{\prime }}\frac{%
\Gamma \left( C\right) \Gamma \left( A+B-C\right) }{\Gamma \left( A\right)
\Gamma \left( B\right) }~.  \label{infinity}
\end{equation}
Now, we choose the exponents $\gamma $ and $\gamma ^{\prime }$ as follows 
\begin{eqnarray}
\gamma  &=&\frac{1}{2}+ \sqrt{ \frac{1}{4} -\left(\omega ^{2}-m^{2}\right) \ell^2 }~,  \label{gammas} \\
\gamma ^{\prime } &=&\frac{1}{2}- \sqrt{ \frac{1}{4} -\left(\omega ^{2}-m^{2}\right) \ell^2 }~.  \notag
\end{eqnarray}
So, imposing the condition that the scalar field be null at spatial
infinity, we can determine the QNFs.
The second term of equation (\ref{infinity}) blows up when $z\rightarrow 1$
unless we impose the condition $A=-n$ or $B=-n$; therefore, we obtain the following set of QNFs:
\begin{equation}
\omega \ell=\frac{i\left(m^2 \ell^2 -n(1+n)\right)}{1+2n}~.
\end{equation}
These QNFs 
are purely imaginary and negative for $m=0$, which guarantees that the Lifshitz black hole is stable under massless scalar field perturbations for the mode with the lowest angular momentum. For $m>0$ there are QNFs with imaginary and positive value, and the Lifshitz black hole is unstable under scalar field perturbations.
Also, we note that  if we interchange the values of the exponents in equation (\ref{gammas})
the same QNFs are obtained. It is worth mentioning that  Eq. (\ref{13}) with $\kappa=0$ can be written as 
\begin{equation}
z(1-z)R''(z)+(1-z)R'(z)+\left(\frac{\omega^2 \ell^2(z y_- - y_+)^2}{(y_+-y_-)^2z(1-z)}-\frac{m^2 \ell^2 }{1-z}\right)R(z)=0~,
\end{equation}
under the change of variable $z=\frac{y-y_+}{y-y_-}$, and if we define $R(z)=z^{\alpha}(1-z)^{\beta}F(z)$, the above equation leads to the hypergeometric equation 
\begin{equation}\label{hypergeometric}
 z(1-z)F''(z)+\left[c-(1+a+b)z\right]F'(z)-ab F(z)=0~,
\end{equation}
 where
\begin{equation}
\alpha= \pm\frac{i\omega \ell y_+}{y_+-y_-}~,
\end{equation}
\begin{equation}
\beta= \frac{1}{2}\left(1\pm \sqrt{1+4 \left( m^2  -\omega^2 \right) \ell^2 }\right)~,
\end{equation}
and the constants are given by
\begin{equation}\label{a}\
a=\alpha+\beta - \frac{i\omega \ell \left|y_-\right|}{(y_+ - y_-)}~,
\end{equation}
\begin{equation}
b=\alpha+\beta + \frac{i\omega \ell \left|y_-\right|}{(y_+ - y_-)}~,
\end{equation}
\begin{equation}
c=1+2\alpha~.
\end{equation}
The general solution of the hypergeometric equation~(\ref{hypergeometric}) is
\begin{equation}
\label{HSolution}
F(z)=c_{1}{_2}F{_1}(a,b,c;z)+c_2z^{1-c}{_2}F{_1}(a-c+1,b-c+1,2-c;z)~,
\end{equation}
and it has three regular singular points at $z=0$, $z=1$, and 
$z=\infty$. ${_2}F{_1}(a,b,c;z)$ is a hypergeometric function
and $c_{1}$ and $c_{2}$ are integration constants. Note that the above QNFs could be computed using the solution (\ref{HSolution}).

\subsubsection{Case $Q=\pm \infty$}
In this case it is possible to obtain an analytical solution for all values of the angular
momentum $\kappa $. Thus, for $\lambda =-\ell^2$ or equivalently $Q=\pm \infty$,
the radial equation (\ref{first}) can be written as 
\begin{equation}\label{equationc1}
 z\left(1-z\right)\partial_{z}^2R(z)+\left(1-z\right)\partial_{z}R(z)+\left[\frac{\omega^2 \ell^2}{z(1-z)}-\frac{m^2 \ell^2 }{1-z}-\kappa\right]R(z)=0~,
\end{equation}
where we have considered $z=1-\ell^2/r^2$. Using the decomposition $R(z)=z^\alpha(1-z)^\beta K(z)$, with
\begin{equation}
\alpha_{\pm}=\pm i \omega \ell~,
\end{equation}
\begin{equation}\label{omega2}\
\beta_{\pm}= \frac{1}{2}\left(1\pm \sqrt{1+4(m^2-\omega^2) \ell^2}\right)~,
\end{equation}
we can write (\ref{equationc1}) as a hypergeometric equation for K
\begin{equation}\label{hypergeometric}
 z(1-z)K''(z)+\left[c-(1+a+b)z\right]K'(z)-ab K(z)=0~,
\end{equation}
where the coefficients are given by
\begin{equation}\label{a}\
a=\alpha+\beta\mp\sqrt{-\kappa}~,
\end{equation}
\begin{equation}
b=\alpha+\beta\pm\sqrt{-\kappa}~,
\end{equation}
\begin{equation}
c=1+2\alpha~.
\end{equation}
The general solution of the hypergeometric equation~(\ref{hypergeometric}) is
\begin{equation}
K=C_{1}{_2}F{_1}(a,b,c;z)+C_2z^{1-c}{_2}F{_1}(a-c+1,b-c+1,2-c;z)~,
\end{equation}
and it has three regular singular points at $z=0$, $z=1$, and 
$z=\infty$. ${_2}F{_1}(a,b,c;z)$ is a hypergeometric function
and $C_{1}$ and $C_{2}$ are constants. Thus, the solution for the
radial function $R(z)$ is
\begin{equation}\label{RV}\
R(z)=C_{1}z^\alpha(1-z)^\beta {_2}F{_1}(a,b,c;z)+C_2z^{-\alpha}(1-z)^\beta
{_2}F{_1}(a-c+1,b-c+1,2-c;z)~.
\end{equation}
So, in the vicinity of the horizon, $z=0$ and using
the property $F(a,b,c,0)=1$, the function $R(z)$ behaves as
\begin{equation}\label{Rhorizon}
R(z)=C_1 e^{\alpha \ln z}+C_2 e^{-\alpha \ln z},
\end{equation}
and the scalar field $\psi$, for $\alpha=\alpha_-$, can be written as follows:
\begin{equation}
\psi\sim C_1 e^{-i\omega \ell (t+ \ln z)}+C_2
e^{-i\omega \ell (t-\ln z)}~,
\end{equation}
in which, the first term represents an ingoing wave and the second an outgoing wave in the black hole. So, by imposing that
 only ingoing waves existing at the horizon, this fixes  $C_2=0$. The radial
solution then becomes
 \begin{equation}\label{horizonsolution}
R(z)=C_1 e^{\alpha \ln z}(1-z)^\beta {_2}F{_1}(a,b,c;z)=C_1e^{-i\omega \ell \ln z}(1-z)^\beta {_2}F{_1}(a,b,c;z)~.
\end{equation}
To implement boundary conditions at infinity ($z=1$), we
apply Kummer's formula
for the hypergeometric function \cite{M. Abramowitz},
\begin{equation}\label{relation}
{_2}F{_1}(a,b,c;z)=\frac{\Gamma(c)\Gamma(c-a-b)}{\Gamma(c-a)\Gamma(c-b)}F_1+(1-z)^{c-a-b}\frac{\Gamma(c)\Gamma(a+b-c)}{\Gamma(a)\Gamma(b)}F_2~,
\end{equation}
where,
\begin{equation}
F_1={_2}F{_1}(a,b,a+b-c,1-z)~, 
\end{equation}
\begin{equation}
F_2={_2}F{_1}(c-a,c-b,c-a-b+1,1-z)~.
\end{equation}
With this expression, the radial function~(\ref{horizonsolution}) reads
\begin{equation}\label{R}\
R(z) = C_1 e^{-i\omega \ell \ln z}(1-z)^\beta\frac{\Gamma(c)\Gamma(c-a-b)}{\Gamma(c-a)\Gamma(c-b)} F_1+C_1 e^{-i\omega \ell  \ln
z}(1-z)^{c-a-b+\beta}\frac{\Gamma(c)\Gamma(a+b-c)}{\Gamma(a)\Gamma(b)}F_2~,
\end{equation}
and at infinity it can be written as 
\begin{equation}\label{R2}\
R_{asymp.}(z) = C_1 (1-z)^\beta\frac{\Gamma(c)\Gamma(c-a-b)}{\Gamma(c-a)\Gamma(c-b)}+C_1 (1-z)^{1-\beta}\frac{\Gamma(c)\Gamma(a+b-c)}{\Gamma(a)\Gamma(b)}~.
\end{equation}
So, for  $\beta_+>1$, the field at infinity vanishes
if $(a)|_{\beta_+}=-n$ or $(b)|_{\beta_+}=-n$ for $n=0,1,2,...$, and for  $\beta_-<0$, the field at infinity vanishes 
if $(c-a)|_{\beta_-}=-n$ or $(c-b)|_{\beta_-}=-n$. Therefore, the QNFs are given by
\begin{equation}\label{w1}
\omega \ell=-i \frac{-m^2 \ell^2 + n + n^2 +\kappa \mp \sqrt{-\kappa} (1+2n)}{
 1 + 2 n \mp 2 \sqrt{-\kappa}}~,
\end{equation}
where $\sqrt{-\kappa}=i\sqrt{l(l+1)}$. This expression can be written as {\footnote{The same QNFs can be obtained by imposing that only outgoing waves exist at spatial infinity.}\label{foot}}
\begin{equation}\label{www}
\omega \ell=\pm \frac{ \sqrt{\kappa} \left( -(1+2n)^2+2(-m^2 \ell^2+n+n^2-\kappa)\right)}{(1+2n)^2+4\kappa}-i \frac{(1+2n)(-m^2 \ell^2+n+n^2+\kappa)}{(1+2n)^2+4\kappa}~.
\end{equation}
Because not all the QNFs have a negative imaginary part we conclude that this black hole is not stable under scalar field perturbations for the case when $Q=\pm \infty$.

\subsubsection{Case $Q=1$}
In the extremal case $\lambda=-2\ell^2$, or equivalently $Q=1$, the radial equation (\ref%
{rad}) reads
\begin{equation}
z\left( z-1\right) R^{\prime \prime }\left( z\right) +2\left( z-1\right)
R^{\prime }\left( z\right) +\left( \frac{\omega ^{2} \ell^2}{z^{3}\left( z-1\right) 
}+\frac{\kappa }{z}-\frac{m^{2} \ell^2}{z\left( z-1\right) }\right) R\left(
z\right) =0~,
\end{equation}
and its solution is given by
\begin{eqnarray}
\nonumber R\left( z\right)  &=&C_{1}e^{\frac{i\omega \ell }{z}}Heun_{C}\left( -2i\omega \ell ,%
\sqrt{1-4\left( \omega ^{2} -m^{2} \right) \ell^2 },1,-2\omega ^{2} \ell^2,-\kappa +\frac{1}{2%
}+2\omega ^{2} \ell^2,\frac{z-1}{z}\right)z^{-\frac{3}{2}-\frac{1}{2}\sqrt{1-4\left( \omega ^{2} -m^{2} \right) \ell^2 }} \times  \\
\nonumber &&\left( z-1\right) ^{\frac{1}{2}+\frac{1}{2}\sqrt{1-4\left( \omega ^{2} -m^{2} \right) \ell^2 }%
} \\
\nonumber &&+ C_{2}e^{\frac{i\omega \ell }{z}}Heun_{C}\left( -2i\omega \ell ,-\sqrt{1-4\left(
\omega ^{2} -m^{2} \right) \ell^2 },1,-2\omega ^{2} \ell^2,-\kappa +\frac{1}{2}+2\omega ^{2} \ell^2,%
\frac{z-1}{z}\right) \times  \\
&&z^{-\frac{3}{2}+\frac{1}{2}\sqrt{1-4\left( \omega ^{2} -m^{2} \right) \ell^2 }%
}\left( z-1\right) ^{\frac{1}{2}-\frac{1}{2}\sqrt{1-4\left( \omega ^{2} -m^{2}\right) \ell^2 }%
}~,
\end{eqnarray}
where $Heun_{C}$ is the confluent Heun function. Thus, when $z\rightarrow 1$, and in order to have a regular scalar field at spatial
infinity, we must set $C_{2}=0$; therefore, the solution reduces to
\begin{eqnarray}
\nonumber R\left( z\right) &=& C_{1}e^{\frac{i\omega \ell }{z}}Heun_{C}\left( -2i\omega \ell  ,\sqrt{%
1-4\left( \omega ^{2} -m^{2} \right) \ell^2},1,-2\omega ^{2} \ell^2,-\kappa +\frac{1}{2}%
+2\omega ^{2} \ell^2,\frac{z-1}{z}\right) z^{-\frac{3}{2}-\frac{1}{2}\sqrt{%
1-4\left( \omega ^{2} -m^{2} \right) \ell^2}}\times \\
&& \left( z-1\right) ^{\frac{1}{2}+\frac{1}{2}\sqrt{%
1-4\left( \omega ^{2} -m^{2}\right) \ell^2}}~,
\end{eqnarray}
where the property $Heun_{C}(a,b,c,d,e,0)=1$  was used \cite{Fiziev}. However, we observe that when $z\rightarrow 1$, the scalar field is
null $R(z) \rightarrow 0$; therefore, there are no QNMs in this
case.

\subsection{Scalar field conformally coupled to gravity}

In this section we calculate the QNMs of the $z=0$ Lifshitz black hole for a test scalar field conformally coupled to gravity. The Klein-Gordon equation for a scalar field non-minimally coupled to curvature is 
\begin{equation}
\frac{1}{\sqrt{-g}}\partial _{\mu }\left( \sqrt{-g}g^{\mu \nu }\partial
_{\nu }\right) \psi-\xi \mathcal{R} \psi =m^{2}\psi ~,  \label{KGNM}
\end{equation}%
where $m$ is the mass of the scalar field $\psi $, $\xi$ is the non-mininal coupling parameter and $\mathcal{R}$ is the Ricci scalar, which reads 
\begin{equation}
\mathcal{R}=\frac{\ell ^4-\lambda^2}{2\ell ^2 r^4}+\frac{4\ell^2-\lambda}{2\ell ^2 r^2}-\frac{3}{2\ell^2}~.
\end{equation}
For a conformally coupled scalar field case we must take $m=0$ and $\xi=1/6$. Now, by means of the following ansatz 
\begin{equation}
\psi =e^{-i\omega t}R(r)Y(\theta ,\phi )~,
\end{equation}%
where $Y(\theta ,\phi )$ is a normalizable harmonic function on the two-sphere
which satisfies Eq. (\ref{nabla}),
the Klein-Gordon equation reduces to
\begin{equation}
\frac{1}{4r}\partial _{r}\left( r^{3}f\left( r\right) \partial _{r}R\right)
+\left( \frac{\omega ^{2} \ell^2}{f\left( r\right) }-\frac{\kappa \ell^2}{r^{2}}%
-m^{2}\ell^2-\xi \ell^2\mathcal{R}\right) R\left( r\right) =0~,  \label{radial}
\end{equation}%
which can be written as a one-dimensional Schr\"{o}dinger equation with an effective potential that vanishes at spatial infinity. Therefore, we will consider only outgoing waves at the asymptotic region as boundary condition. It is worth to mention that Eq. (\ref{radial}) only has analytical solution for  $\lambda=-1$ as we will show below. Therefore, we will perform numerical studies for $\lambda\neq -1$ by using the improved AIM \cite{Cho:2009cj}, which is an improved version of the method proposed in Refs. \cite{Ciftci, Ciftci:2005xn} and it has been applied successful in the context of QNMs for different black holes geometries, see for instance \cite{Cho:2009cj, Cho:2011sf, Catalan:2013eza, Zhang:2015jda, Barakat:2006ki}.

\subsubsection{Numerical analysis}
In order to implement the improved AIM we make the following consecutive change of variables $y=r^2$ and $z=\frac{y-y+}{y-y-}$ to Eq. (\ref{KGNM}), as we do in the previous sections. Then, the Klein-Gordon equation yields
\begin{eqnarray}
\nonumber && z(1-z )R''(z)+(1-z)R'(z)\\
&&+\left( \frac{\omega^2 \ell^2 (zy_--y_+)^2}{(y_+-y_-)^2z(1-z)}+\frac{\kappa \ell^2}{zy_--y_+}-\frac{\xi (\ell^4-\lambda^2)(1-z)}{2 (zy_--y_+)^2}+\frac{\xi(4\ell^2-\lambda)}{2(zy_--y_+)}+\frac{3 \xi}{2(1-z)}\right)R(z)=0~. \label{numericalmethod}
\end{eqnarray}
Now, we must consider the behavior of the scalar field on the event horizon and at spatial infinity. Accordingly, on the horizon, $z\rightarrow 0$, the behavior of the scalar field is given by 
\begin{equation}
R\left( z\rightarrow 0\right) \sim C_{1}z^{-\frac{i\omega \ell y_+}{y_+-y_-}}+C_{2}y^{\frac{i\omega \ell y_+}{y_+-y_-}}~,
\end{equation}
So, if we consider only ingoing waves on the horizon, we must impose $C_{2}=0$. 
Asymptotically, from Eq. (\ref{numericalmethod}), the scalar field behaves as 
\begin{equation}
R \left( z\rightarrow 1\right) \sim D_{1}\left( 1-z\right)
^{1/2-i \omega \ell}+D_{2}\left( 1-z\right) ^{1/2+i \omega \ell}~.
\end{equation}
So, in order to have only outgoing waves at infinity we must impose $D_{2}=0 $. Therefore, taking into account these behaviors we define 
\begin{equation}
R\left( z\right) =z^{-\frac{i\omega \ell y_+}{y_+-y_-}} \left( 1-z\right)
^{1/2-i \omega \ell} \chi \left(
z\right)~.
\end{equation}
Then, by inserting these fields in Eq. (\ref{numericalmethod}) we obtain the homogeneous
linear second-order differential equation for the function $\chi (z)$ 
\begin{equation}
\chi ^{\prime \prime }=\lambda _{0}(z)\chi ^{\prime }+s_{0}(z)\chi ~,
\label{de}
\end{equation}%
where
\begin{eqnarray}
\lambda _{0}(z) &=& -\frac{(y_+-y_-)(1-2z)-2i \omega \ell(y_++y_-z-2y_+z)}{(y_+-y_-)z(1-z)}~, \\
s_{0}(z) &=& \frac{\ell^4+3y_+^2+z(-\ell^4+y_- (-4\ell^2+3y_-z-12\kappa \ell^2))+\lambda y_-z-\lambda^2 (1-z) -y_- (-4\ell^2+6y_-z-12 \kappa \ell^2 +\lambda)}{12z(1-z)(y_+-y_-z)^2} \\ \notag
 &+& \frac{12 i \omega \ell (y_--2y_+)(y_+-y_-z)^2-48 \omega^2 \ell^2 y_+(y_+-y_-z)^2}{12z(1-z)(y_+-y_-)(y_+-y_-z)^2}~.
\end{eqnarray}
Then, in order to implement the improved AIM it is necessary to differentiate Eq. (\ref{de}) $n$ times with respect to $z$,
which yields the following equation: 
\begin{equation}
\chi ^{n+2}=\lambda _{n}(z)\chi ^{\prime }+s_{n}(z)\chi~,  \label{de1}
\end{equation}%
where 
\begin{equation}
\lambda _{n}(z)=\lambda _{n-1}^{\prime }(z)+s_{n-1}(z)+\lambda
_{0}(z)\lambda _{n-1}(z)~,  \label{Ln}
\end{equation}%
\begin{equation}
s_{n}(z)=s_{n-1}^{\prime }(z)+s_{0}(z)\lambda _{n-1}(z)~.  \label{Snn}
\end{equation}%
Then, by expanding the $\lambda _{n}$ and $s_{n}$ in a Taylor series around
the point $\delta $, at which the improved AIM is performed 
\begin{equation}
\lambda _{n}(\delta )=\sum_{i=0}^{\infty }c_{n}^{i}(z-\delta )^{i}~,
\end{equation}%
\begin{equation}
s_{n}(\delta )=\sum_{i=0}^{\infty }d_{n}^{i}(z-\delta )^{i}~,
\end{equation}%
where the $c_{n}^{i}$ and $d_{n}^{i}$ are the $i^{th}$ Taylor coefficients
of $\lambda _{n}(\delta )$ and $s_{n}(\delta )$, respectively, and by replacing
the above expansions in Eqs. (\ref{Ln}) and (\ref{Snn}) the following
set of recursion relations for the coefficients is obtained:
\begin{equation}
c_{n}^{i}=(i+1)c_{n-1}^{i+1}+d_{n-1}^{i}+%
\sum_{k=0}^{i}c_{0}^{k}c_{n-1}^{i-k}~,
\end{equation}%
\begin{equation}
d_{n}^{i}=(i+1)d_{n-1}^{i+1}+\sum_{k=0}^{i}d_{0}^{k}c_{n-1}^{i-k}~.
\end{equation}%
In this manner, the authors of the improved AIM have avoided the
derivatives that contain the AIM in \cite{Cho:2009cj, Cho:2011sf}, and
the quantization condition, which is equivalent to imposing a termination
to the number of iterations, is given by 
\begin{equation}
d_{n}^{0}c_{n-1}^{0}-d_{n-1}^{0}c_{n}^{0}=0~.
\end{equation}
We solve this equation numerically to find the QNFs. In Table \ref{QNM1}, we show some lowest QNFs, for a scalar field conformally coupled to curvature with $\kappa= 0$,  $\ell=1$ and different values of $\lambda$. Then, in Table \ref{QNM2}, we show some fundamentals QNFs, for a scalar field conformally coupled to curvature with $\kappa= 0, 1, 2, 3$,  $\ell=1$ and different values of $\lambda$. We observe that the QNFs have real and imaginary parts, with an imaginary part that is negative, which ensures the stability of the $4$-dimensional Lifshitz black hole under scalar perturbations. 
\begin{table}[ht]
\caption{Quasinormal frequencies for $\protect\kappa= 0$, $\ell=1$ and different values of $\lambda$.}
\label{QNM1}\centering
\begin{tabular}{cccccc}
\hline\hline
$n$ & $\lambda=-1.3 $ & $\lambda=-1$ & $\lambda=-0.7 $ \\[0.5ex] \hline
$0$ & $0.29170-0.22801i$ &  $0.32275-0.25000i$ & $0.35757-0.27902i$ \\ 
$1$ & $0.27600-0.68476i$ &  $0.32275-0.75000i$ & $0.34211-0.83763i$ \\ 
$2$ & $0.23817-1.14518i$ &  $0.32275-1.25000i$ & $0.30647-1.39878i$ \\ 
$3$ & $0.13120-1.63955i$ &  $0.32275-1.75000i$ & $0.22919-1.97158i$ \\[0.5ex] \hline
\end{tabular}%
\end{table}

\begin{table}[ht]
\caption{Fundamentals Quasinormal frequencies ($n= 0$), $\ell=1$ and different values of $\kappa$ and $\lambda$.}
\label{QNM2}\centering
\begin{tabular}{cccccc}
\hline\hline
$\kappa$ & $\lambda=-1.9 $ & $\lambda=-1.3 $ & $\lambda=-1$ & $\lambda=-0.7 $ & $ \lambda=-0.3 $ \\[0.5ex] \hline
$0$ & $0.22840-0.20893i$ & $0.29170-0.22801i$ & $0.32275-0.25000i$ & $0.35757-0.27902i$ & $0.41878-0.33808i$ \\
$2$ & $0.60092-0.19838i$ & $0.70778-0.22800i$ & $0.77728-0.25000i$ & $0.86642-0.27902i$ & $1.04077-0.33703i$ \\
$6$ & $0.98966-0.19724i$ & $1.15443-0.22800i$ & $1.26656-0.25000i$ & $1.41292-0.27902i$ & $1.70231-0.33686i$ \\
$12$ & $1.38091-0.19692i$ & $1.60635-0.22800i$ & $1.76186-0.25000i$ & $1.96592-0.27902i$ & $2.37061-0.33682i$ \\[0.5ex] \hline\end{tabular}%
\end{table}

\subsubsection{Case $Q=\pm \infty$}
In this case ($\lambda=-\ell^2$), due to the simplicity of the Ricci scalar, it is possible to obtain an analytical solution. The radial equation (\ref{radial}) reads
\begin{equation}\label{nn}
\frac{1}{4r}\partial _{r}\left( r^{3}f\left( r\right) \partial _{r}R\right)
+\left( \frac{\omega ^{2} \ell^2}{f\left( r\right) }-\frac{\kappa \ell^2+5 \xi \ell^2 /2 }{r^{2}}%
-m^{2}\ell^2+\frac{3 \xi}{2}\right) R\left( r\right) =0~.
\end{equation}
So, if we compare this equation with the analogous equation of the mininal case (Eq. (\ref{first})), we see that is possible to obtain (\ref{nn}) by means of the following substitutions in Eq. (\ref{first}):
\begin{eqnarray}\label{sust}
\notag\kappa & \rightarrow & \kappa+\frac{5 \xi}{2}~, \\
m^2 & \rightarrow & m^2-\frac{3 \xi}{2 \ell^2}~.
\end{eqnarray} 
Thus, using the above substitutions in the QNFs (\ref{www}) we find
\begin{equation}
\omega \ell=\pm \frac{ \sqrt{\kappa+\frac{5 \xi}{2}} \left( -(1+2n)^2+2(-m^2 \ell^2+n+n^2-\kappa-\xi)\right)}{(1+2n)^2+4\kappa+10\xi}-i \frac{(1+2n)(-m^2 \ell^2+n+n^2+\kappa+4\xi)}{(1+2n)^2+4\kappa+10\xi}~,
\end{equation}
and for a conformal scalar field ($m=0$, $\xi=1/6$) this equation yields
\begin{equation}
\omega \ell=\pm \frac{ \sqrt{\kappa+\frac{5}{12}} \left( -(1+2n)^2+2(n+n^2-\kappa-\frac{1}{6})\right)}{(1+2n)^2+4\kappa+\frac{5}{3}}-i \frac{(1+2n)(n+n^2+\kappa+\frac{2}{3})}{(1+2n)^2+4\kappa+\frac{5}{3}}~,
\end{equation}
which satisfies that there are only outgoing waves at the asymptotic region, see footnote $1$. 
Clearly the imaginary part of the QNFs is negative, which ensures the stability of this black hole under conformally coupled scalar field perturbations. These QNFs agrees with the numerical results for $\lambda=-1$, $\kappa=0$ and $\ell=1$ showed in Table (\ref{QNM1}).

\section{Reflection and transmission coefficients and absorption cross section}
In this section, we focus our attention to the minimally coupled scalar field case. However, a similar analysis can be performed for scalar fields conformally coupled to gravity, and for $\lambda=-\ell^2$ the results are straightforward obtained from the minimal case by using the substitutions (\ref{sust}) and taking $m=0$, $\xi=1/6$. 
The reflection and the transmission coefficients are defined by
\begin{equation}\label{reflectiond}\
\Re :=\left|\frac{F_{\mbox{\tiny asymp}}^{\mbox{\tiny
out}}}{F_{\mbox{\tiny asymp}}^{\mbox{\tiny in}}}\right|, \qquad
\mbox{and} \qquad  \mathfrak{T}:=\left|\frac{F_{\mbox{\tiny
hor}}^{\mbox{\tiny in}}}{F_{\mbox{\tiny asymp}}^{\mbox{\tiny
in}}}\right|,
\end{equation}
where $F$ is the flux given by
\begin{equation}\label{fluxd}\
\textit{F}=\frac{1}{2i}\sqrt{-g}g^{rr}\left(R^{\ast}\partial_{r}R-R\partial_{r}R^{\ast}\right).
\end{equation}
So, in order to calculate the above coefficients we need to know the behavior
of the radial function both on the horizon and at asymptotic
infinity.

\subsection{Case $\kappa = 0$}

In this case, the behavior at the horizon is given by Eq. (\ref{HSolution}) with $c_2=0$, and choosing the negative value of $\alpha$ and using Eq. (\ref{fluxd}), we get the following flux on the horizon:
\begin{equation}
\textit{F}
_{hor}^{in}=-\omega \sin \theta \left|c_{1}\right|^{2} y_+~.
\end{equation}
On the other hand, by applying Kummer's formula (\ref{relation})  for the hypergeometric function in Eq. (\ref{HSolution}), the asymptotic behavior of $R(z)$  can be written as
\begin{equation}
R\left( z \rightarrow 1 \right) = \bar{c}_1 (z-1)^{\beta}+\bar{c}_2 (z-1)^{1-\beta},
\end{equation}
where
\begin{eqnarray}
\bar{c}_1& = & c_{1}\frac{\Gamma \left( c\right) \Gamma \left( c-a-b\right) }{\Gamma \left(
c-a\right) \Gamma \left( c-b\right) }~, \notag \\
\bar{c}_2& = & c_{1}\frac{\Gamma \left( c\right) \Gamma \left( a+b-c\right) }{\Gamma \left( a\right) \Gamma \left( b\right) }~.
\end{eqnarray}
Thus, using Eq. (\ref{fluxd}) we obtain the flux
\begin{equation}
\textit{F}_{asymp.}=\frac{\sin \theta}{\ell} (y_{+}-y_-) \sqrt{ \left( \omega^2 -m^2 \right) \ell^2-\frac{1}{4}}\left(\left|\Bar{c}_2\right|^{2}-\left|\Bar{c}_1\right|^{2}\right)~,
\end{equation}
for $\beta=\beta_{+}$, 
the reflection and transmission coefficients are given by
\begin{equation}
\label{RE}
\Re=\frac{\left|\bar{c}_2\right|^2}{\left|\bar{c}_1\right|^2}~,
\end{equation}
\begin{equation}
\label{TE}
\mathfrak{T}=\frac{\omega \ell\left| c_{1}\right|^2 y_+}{(y_+ - y_-)\left|\bar{c}_1\right|^2\sqrt{ \left( \omega^2 -m^2 \right) \ell^2-\frac{1}{4}}}~,
\end{equation}
and the absorption cross section, $\sigma_{abs}$, is given by
\begin{equation}\label{absorptioncrosssection}\
\sigma_{abs}=\frac{\ell \left| c_{1}\right|^2 y_+}{(y_+ - y_-)\left|\bar{c}_1\right|^2\sqrt{\left( \omega^2 -m^2 \right) \ell^2-\frac{1}{4}}}~.
\end{equation}

\begin{figure}
\includegraphics[width=4.0in,angle=0,clip=true]{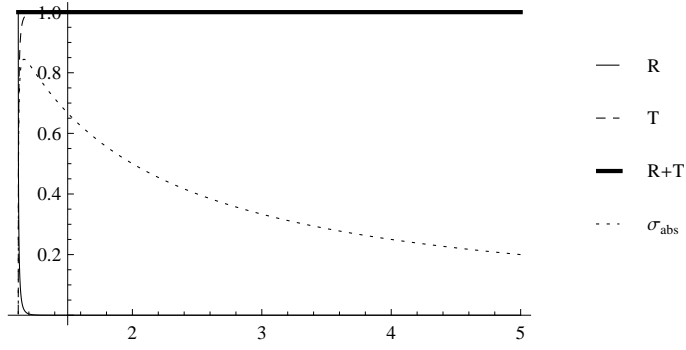}
\caption{The reflection coefficient $R$ (solid curve), the transmission coefficient $T$ (dashed curve), $R+T$ (thick curve) and the absorption cross section $\sigma_{abs}$ (dotted curve) as a function of $\omega$, for $m=1$, $\ell=1$, and $\lambda=-1.9$.}
\label{coeffB}
\end{figure}

\begin{figure}
\includegraphics[width=4.5in,angle=0,clip=true]{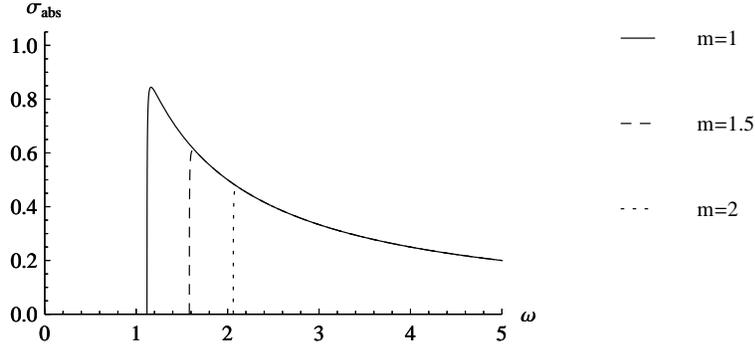}
\caption{The behavior of $\sigma_{abs}$ as a function of $\omega$, for $\lambda=-1.9$, $\ell=1$, and $m = 1, 1.5, 2$.}
\label{SigmaB}
\end{figure}

Interestingly, the poles of the transmission coefficient yields the same set of QNFs found in the previous section, which is equivalent to imposing as a boundary condition that only outgoing waves exist at asymptotic infinity. Now, we perform a numerical analysis of the reflection coefficient~(\ref{RE}), transmission coefficient~(\ref{TE}) and absorption cross section~(\ref{absorptioncrosssection}) of the four-dimensional Lifshitz black hole with $z=0$ for scalar fields. So, we plot the reflection and transmission coefficients and the absorption cross section in Fig.~(\ref{coeffB}) for scalar fields with $m=1$. Essentially, we found that the reflection coefficient is 1 at low frequency limit, and for high frequency limit this coefficient is null, with the behavior of the transmission coefficient being opposite with $R+T = 1$. In addition, the absorption cross section is null in the low and high-frequency limit, but there is a range of frequencies for which the absorption cross section is not null, and it also has a maximum value in the low-frequency limit (see Fig.~(\ref{SigmaB})). Furthermore, we observe that the absorption cross section can take higher values when the mass of the scalar field decreases (Fig.~(\ref{SigmaB})) in the low frequency limit. However,
beyond a certain value of the frequency the absorption cross section does not depend on the mass of the scalar field.


\subsection{Case $Q=\pm \infty$}
\begin{figure}
\includegraphics[width=4.0in,angle=0,clip=true]{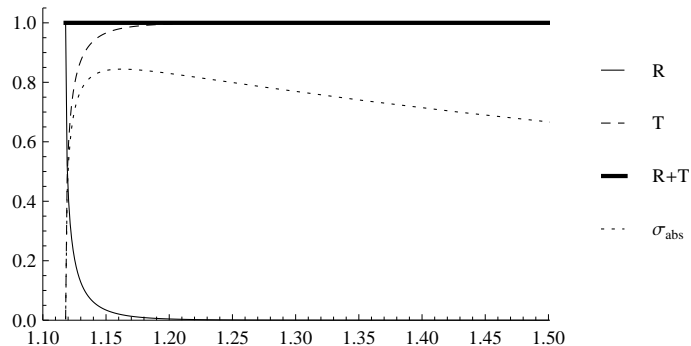}
\caption{The reflection coefficient $R$ (solid curve), the transmission coefficient $T$ (dashed curve), $R+T$ (thick curve) and the absorption cross section $\sigma_{abs}$ (dotted curve) as a function of $\omega$, for $m=1$, $\ell=1$ and $\kappa=0$.}
\label{coeffA}
\end{figure}

\begin{figure}
\includegraphics[width=4.5in,angle=0,clip=true]{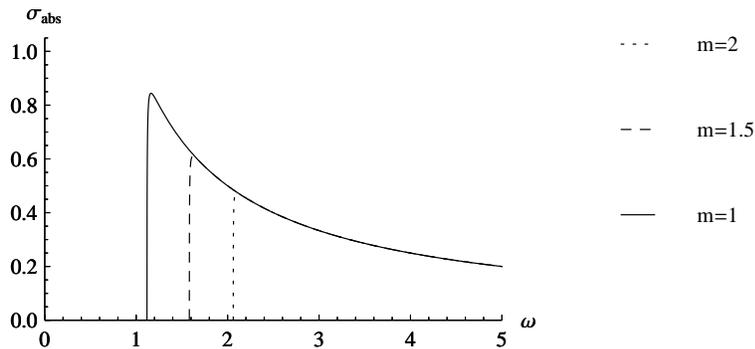}
\caption{The behavior of $\sigma_{abs}$ as a function of $\omega$, for $\kappa=0$, $\ell=1$ and $m = 1, 1.5, 2$.}
\label{SigmaA}
\end{figure}

In this case the behavior at the horizon is given by Eq. (\ref{Rhorizon}), with $C_2=0$, and using Eq. (\ref{fluxd}), we get the
flux at the horizon
\begin{equation}
\textit{F}
_{hor}^{in}=-\omega \ell^2 \sin \theta \left|C_{1}\right|^{2}~.
\end{equation}
On the other hand, the asymptotic behavior of $R(z)$ is given by Eq. (\ref{R2}), which can be written as
\begin{equation}
R\left( z \rightarrow 1 \right) = B_{1} (z-1)^{\beta}+B_{2} (z-1)^{1-\beta},
\end{equation}
where
\begin{eqnarray}
B_{1} & = & C_{1}\frac{\Gamma \left( c\right) \Gamma \left( c-a-b\right) }{\Gamma \left(
c-a\right) \Gamma \left( c-b\right) }~, \notag \\
B_{2} & = & C_{1}\frac{\Gamma \left( c\right) \Gamma \left( a+b-c\right) }{\Gamma \left( a\right) \Gamma \left( b\right) }~.
\end{eqnarray}
Thus, using Eq. (\ref{fluxd}), we get the flux
\begin{equation}\label{fluxdinfinity}\
\textit{F}_{asymp.}=\ell \sin \theta \sqrt{ \left( \omega^2 -m^2 \right) \ell^2-\frac{1}{4}}\left(\left| B_2\right|^{2}-\left| B_1\right|^{2}\right)
\end{equation}
Therefore, the reflection and transmission coefficients
 are given by
\begin{equation}\label{REA}
\Re=\frac{\left| B_2\right|^2}{\left| B_1\right|^2}~,
\end{equation}
\begin{equation}\label{TEA}
\mathfrak{T}=\frac{\omega \ell\left|C_{1}\right|^2}{\sqrt{ \left( \omega^2 -m^2 \right) \ell^2-\frac{1}{4}}\left| B_1\right|^2}~,
\end{equation}
and the absorption cross section, $\sigma_{abs}$, is given by
\begin{equation}\label{absorptioncrosssectionA}\
\sigma_{abs}=\frac{\mathfrak{T}}{\omega}=\frac{\ell \left|C_{1}\right|^2}{\sqrt{ \left( \omega^2 -m^2 \right) \ell^2-\frac{1}{4}}\left| B_1\right|^2}~.
\end{equation}
As in the previous case, the poles of the transmission coefficient yields the same set of QNFs found in the previous section. Also, 
we observe the same behavior described in the previous case for the reflection coefficient~(\ref{REA}), transmission coefficient~(\ref{TEA}), and absorption cross section~(\ref{absorptioncrosssectionA}), i.e., 
we have found that the reflection coefficient is 1 at the low frequency limit, and for the high frequency limit this coefficient is null, with the behavior of the transmission coefficient being  opposite with $R+T = 1$ (see  Fig.~(\ref{coeffA})). Also, the absorption cross section is null in the low and high frequency limits, but there is a range of frequencies for which the absorption cross section is not null, and it also has a maximum value in the low frequency limit (see Fig.~(\ref{SigmaA})). Furthermore, we observe that the absorption cross section can take higher values when the mass of the scalar field decreases (Fig.~(\ref{SigmaA})) in the low frequency limit. However,
beyond a certain value of the frequency the absorption cross section does not depend on the mass of the scalar field. It is worth noting that the absorption cross section does not depend on the angular momentum of the scalar field being the same for every value of $\kappa$.

\section{Conclusions}
In this work we calculated the QNFs of scalar field perturbations for the
four-dimensional asymptotically Lifshitz black hole in conformal gravity with a spherical topology and dynamical exponent $z=0$, where the anisotropic scale invariance corresponds to a space-like scale invariance with no transformation of time, for some special cases that depend on $Q$, and by imposing suitable boundary conditions at spatial infinity. These scalar fields are considered as mere test fields, without backreaction over the spacetime itself. Therefore, it is not necessary for such fields to have the same symmetries as the background spacetime. However, if one considers the backreaction of the matter fields over the spacetime, the relation between the symmetries of the spacetime and the matter fields is not trivial. For conformal gravity the gravitational field equations  imply that the trace of the stress-energy tensor must vanish due to the Bach tensor is traceless.

Firstly, we analyzed massive scalar field perturbations minimally coupled to curvature, which does not have the same symmetries as the background spacetime due to the trace of stress-energy tensor is not null. The first case studied corresponds to a scalar field without angular momentum ($\kappa=0$), and we found that there is a spectrum of quasinormal frequencies for which the scalar field becomes null at spatial infinity. These frequencies are purely imaginary and negative for $m=0$;
however, for $m \neq 0$ some QNFs are imaginary and positive.
Another case we analyzed corresponds to $Q=\pm \infty$, where we found a spectrum of QNFs that respect the Dirichlet boundary condition; however, some of them have a positive imaginary part. Therefore, the black hole is unstable under massive scalar field perturbations and stable under massless scalar field perturbations.
Also, we analyzed the extremal case $Q=1$, for which we found that there are no QNMs as in \cite{Crisostomo:2004hj}, where the authors demonstrated the absence of QNMs in the extremal BTZ black hole. However, it was shown that it is possible to construct the QNMs of three-dimensional extremal black holes algebraically as the descendants of the highest weight modes \cite{Chen:2010sn}, with  hidden conformal symmetry being an intrinsic property of the extremal black hole. Also, it is worth mentioning that QNMs for extremal black holes are not always absent, for instance see \cite{Afshar:2010ii}, where the authors reported the presence of QNMs for extremal BTZ black holes in topologically massive gravity. 

On the other hand, because the gravitational field equations imply that the trace of the stress-energy tensor must vanish, we also considered scalar field perturbations conformally coupled to curvature  which have a traceless stress-energy tensor, and we showed that the imaginary part of the QNFs calculated is negative, what guaranties the stability of the Lifshitz black hole under conformally coupled scalar field perturbations, this was shown by using the improved AIM and analytical solutions. This behavior is similar to the studied in \cite{Catalan:2014una} for a three-dimensional Lifshitz black hole in conformal gravity. 

Finally, we focused our attention to the minimally coupled scalar field case and we computed the reflection and transmission coefficients and the absorption cross section, and we showed numerically that the absorption cross section vanishes at the low and high frequency limits. Therefore, a wave emitted from the horizon, with low or high frequency, does not reach the spatial infinity and is totally reflected, because the fraction of particles penetrating the potential barrier vanishes. However, we have shown there is a range of frequencies where the absorption cross section is not null. The reflection coefficient is 1 at the low frequency limit and for the high frequency limit this coefficient is null, with the behavior of the transmission coefficient being opposite with $R+T=1$. 
Also, we have shown that the absorption cross section increases if the mass of the scalar field decreases in the low frequency limit; however, beyond a certain value of the frequency the absorption cross section does not depend on the mass of the scalar field. Furthermore, we have shown that the absorption cross section does not depend on the angular momentum.
\section*{Acknowledgments}
The authors would like to thank the referee for useful comments and constructive suggestions. This work was funded by the Comisi{\'o}n Nacional de Investigaci{\'o}n Cient{%
\'i}fica y Tecnol{\'o}gica through FONDECYT Grants 11140674 (PAG), 11121148 (YV, MC) and also by the Direcci{\'o}n de Investigaci{\'o}n, Universidad de La
Frontera (EC, MC). P. A. G. acknowledges the hospitality of the Universidad de La Serena where part of this work was undertaken.  P. A. G. and Y. V. acknowledge the hospitality of the National Technical University of Athens. 
\appendix

\end{document}